\documentclass[prb,twocolumn,preprintnumbers,amsmath,amssymb,longbibliography]{revtex4}
\usepackage{epsfig}
\usepackage{graphicx}
\usepackage{dcolumn}
\usepackage{color}
\usepackage{natbib}
\usepackage{bm}

\begin{document}

\title{Gaps tunable by electrostatic gates in strained graphene}
\author{T. Low$^{1,2}$}
\author{F. Guinea$^3$}
\author{M. I. Katsnelson$^4$}
\affiliation{$^1$ IBM T.J. Watson Research Center, Yorktown Heights, NY 10598, USA\\
$^2$ Network for Computational Nanoelectronics, Purdue University, West Lafayette, IN 47907, USA \\
$^3$ Instituto de Ciencia de Materiales de Madrid. CSIC. Sor Juana In\'es de la Cruz 3. 28049 Madrid. Spain \\
$^4$ Radboud University Nijmegen, Institute for Molecules and Materials,
Heyendaalseweg 135, 6525AJ Nijmegen, The Netherlands}

\begin{abstract} We show that when the pseudomagnetic fields created by long wavelength deformations
are appropriately coupled with a scalar electric potential, a
significant energy gap can emerge due to the formation of a
Haldane state.  Ramifications of this physical effect are examined through
the study of various strain geometries commonly seen in experiments, such as strain superlattices and wrinkled suspended graphene. Of particular technological importance, we consider setup where this gap can be tunable through electrostatic gates, allowing for the design of electronic devices
not realizable with other materials.
\end{abstract}

 \maketitle

\section{Introduction}
 Graphene\cite{Netal04,Netal05} is a material whose unique
properties are a fascinating challenge in both fundamental and applied sciences.
The basic properties of its electronic structure
are chirality, electron-hole symmetry, and linear gapless energy
spectrum, that is, charge carriers in graphene are massless Dirac
fermions\cite{NGPNG09}. In addition, corrugations and topological defects
create gauge (pseudomagnetic and even pseudo-gravitational) fields
acting on electron states\cite{VKG10}.

Recently, a novel state of matter, a quantum Hall insulator
without a macroscopic magnetic field (Haldane state\cite{H88}),
has spawned the interest in unusual topological properties of band
structures, leading to the prediction of topological insulators in
two and three dimensions\cite{KM05b,FKM07,QZ10,M10,HK10}. It was
understood afterwards that such Haldane state can be realized in a
graphene superlattice by a suitable combination of scalar and
vector electromagnetic potentials\cite{S09}. A gap opens in the
electronic spectrum, turning graphene into a quantum Hall
insulator with protected chiral edge states. Since, long
wavelength strains in graphene induce a pseudomagnetic gauge
field\cite{noteMGV07,GKG10}, the combination of strains and a
scalar potential should too open a gap in graphene\footnote{In
contrast with the real magnetic field, strains do not break time
reversal symmetry, and the resulting insulator is not a strong
topological insulator - there is a {\it pair} of
counterpropagating edge states belonging to different valleys,
however, the intravalley scattering is frequently very weak which
make these states well protected, similar to the case considered
in Ref.\cite{GKG10}. Although strong edge disorder would leads to
a transport gap instead\cite{LG10}, which might be technologically
useful too. }. The latter, if realised, would be of general
interest. This is the subject of our work.

Energy gap in graphene is crucial for many applications. Its
realization remains a challenging problem, as the transformation
of electrons into holes i.e. Klein tunneling\cite{KNG06}, is an
significant obstacle. At present, all known ways of gap opening
have a detrimental effect on the electron mobility. In both biased
bilayer and chemically functionalized graphene, one arrives at a
disordered semiconductor with Mott variable range hopping
mobility~\cite{TJH10,ENM09,NRJR10}. In graphene nanoribbons, the
mobility is typically several order of magnitude smaller than in
bulk graphene~\cite{HBK10}. Our approach provides an attractive
route to circumvent these limitations and might also allows for
the design of new electronic devices.

\begin{figure}
\scalebox{0.6}[0.6]{\includegraphics*[viewport=200 45 780 570]{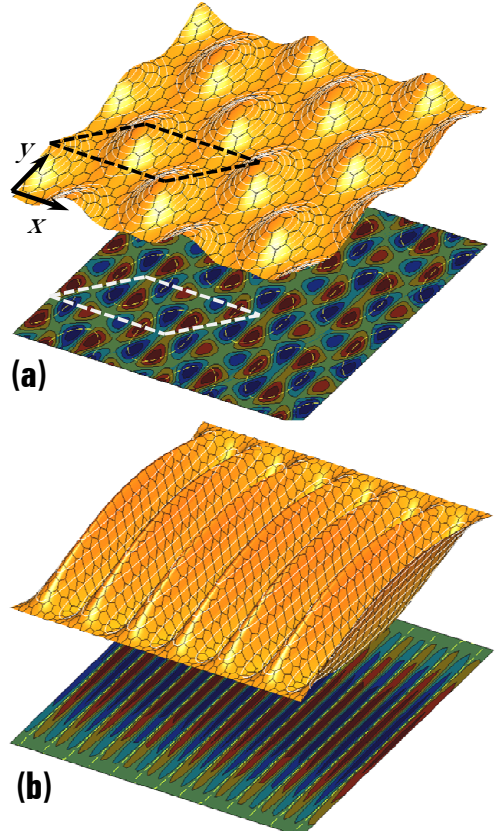}}
\caption[fig]{\textcolor{black}{Two strained graphene lattice configurations (upper panel) and their corresponding pseudomagnetic field (lower panel) are studied in this paper i.e. $\bold{(a)}$ strain superlattice and $\bold{(b)}$ wrinkled graphene. } }
 \label{sketch}
\end{figure}

This paper is organized as follows. In Sec. II, we present a
general argument for the gap opening due to combination of strains
and a scalar potential. Sec. III considers specific realization of
this effect in strain superlattices and wrinkles,
Fig.~\ref{sketch} provide an illustrations of these two strain
geometries. In particular, we consider physical setup where this
gap can be tunable through electrostatic gates. Sec. IV considers
related effects such as Fermi velocity renormalization,
topological defects and interplay with magnetic field.

\section{\label{gen} General arguments}

Strain induces scalar and gauge potentials in graphene\cite{VKG10}. Fig.~\ref{sketch} illustrates the strain induced pseudo magnetic field for a strain superlattice and wrinkled graphene. In terms of the strain tensor, the scalar and vector potentials are\cite{SA02b,M07,VKG10}:
\begin{align}
V ( \vec{\bf r} ) &= g \left[ u_{xx} ( \vec{\bf r} ) + u_{yy} ( \vec{\bf r} ) \right] \nonumber \\
A_x ( \vec{\bf r} ) &= \frac{\beta}{a} \left[ u_{xx} ( \vec{\bf r} ) - u_{yy} ( \vec{\bf r} ) \right] \nonumber \\
A_y ( \vec{\bf r} ) &= 2 \frac{\beta}{a} u_{xy} ( \vec{\bf r} )
\label{fields}
\end{align}
where $g \approx 4$eV\cite{CJS10}, $\beta = - \partial \log ( t )
/
\partial \log ( a ) \approx 2$\cite{HKSS88}, $t \approx 3$eV is
the hopping between $\pi$ orbitals in nearest neighbor carbon
atoms, and $a \approx 1.42$\, \AA \, is the distance between
nearest neighbor atoms.

The Hamiltonian for Dirac fermions in graphene, including a scalar
potential and gauge fields due to strains is given by
\begin{align}
{\cal H} &= {\cal H}_0 + {\cal H}_A + {\cal H}_V \nonumber \\
{\cal H}_0 &= v_F \left( i \sigma_x \tau_z \partial_x  + i \sigma_y \partial_y  \right)  \nonumber \\
{\cal H}_A &= - v_F \left[ \sigma_x A_x ( \vec{\bf r} ) + \tau_z \sigma_y A_y ( \vec{\bf r} ) \right] \nonumber \\
{\cal H}_V &=  V ( \vec{\bf r} )
\label{hamil}
\end{align}
where $v_F$ is the Fermi velocity, and $\sigma_{i=x,y,z}$ and
$\tau_{i=x,y,z}$ are Pauli matrices which operates on the
sublattice and valley indices.

 Using perturbation theory in both
scalar and vector potential, we obtain a self energy correction
which can be written as a cross-term to the Hamiltonian (linear in
$V$ and linear in ${\bf A}$):
\begin{eqnarray}
\nonumber
\Sigma ( \omega , \vec{\bf r} - \vec{\bf r}' )
&=& {\cal H}_V
\left( \omega {\cal I} - {\cal H}_0 \right)^{-1} {\cal H}_A +\\
\nonumber
&&{\cal H}_A \left( \omega {\cal I} - {\cal H}_0 \right)^{-1} {\cal
H}_V  \\
\nonumber
 &=& {\cal H}_V \left( \omega {\cal I} + {\cal H}_0 \right)
\left( \omega^2 {\cal I}^2 - {\cal H}_0^2 \right)^{-1} {\cal H}_A+\\
 && \left( {\cal H}_V \leftrightarrow {\cal H}_A \right)
\label{sigma}
\end{eqnarray}
where the identity matrix ${\cal I}$ is made explicit. Since the
self energy contains term proportional to $\sigma_x \sigma_y =
i\sigma_z$, a gap can be opened in both valleys. Next we make
explicit the necessary relations between the scalar and gauge
potential for this gap opening.

\begin{figure}
\scalebox{0.5}[0.5]{\includegraphics*[viewport=150 70 725 565]{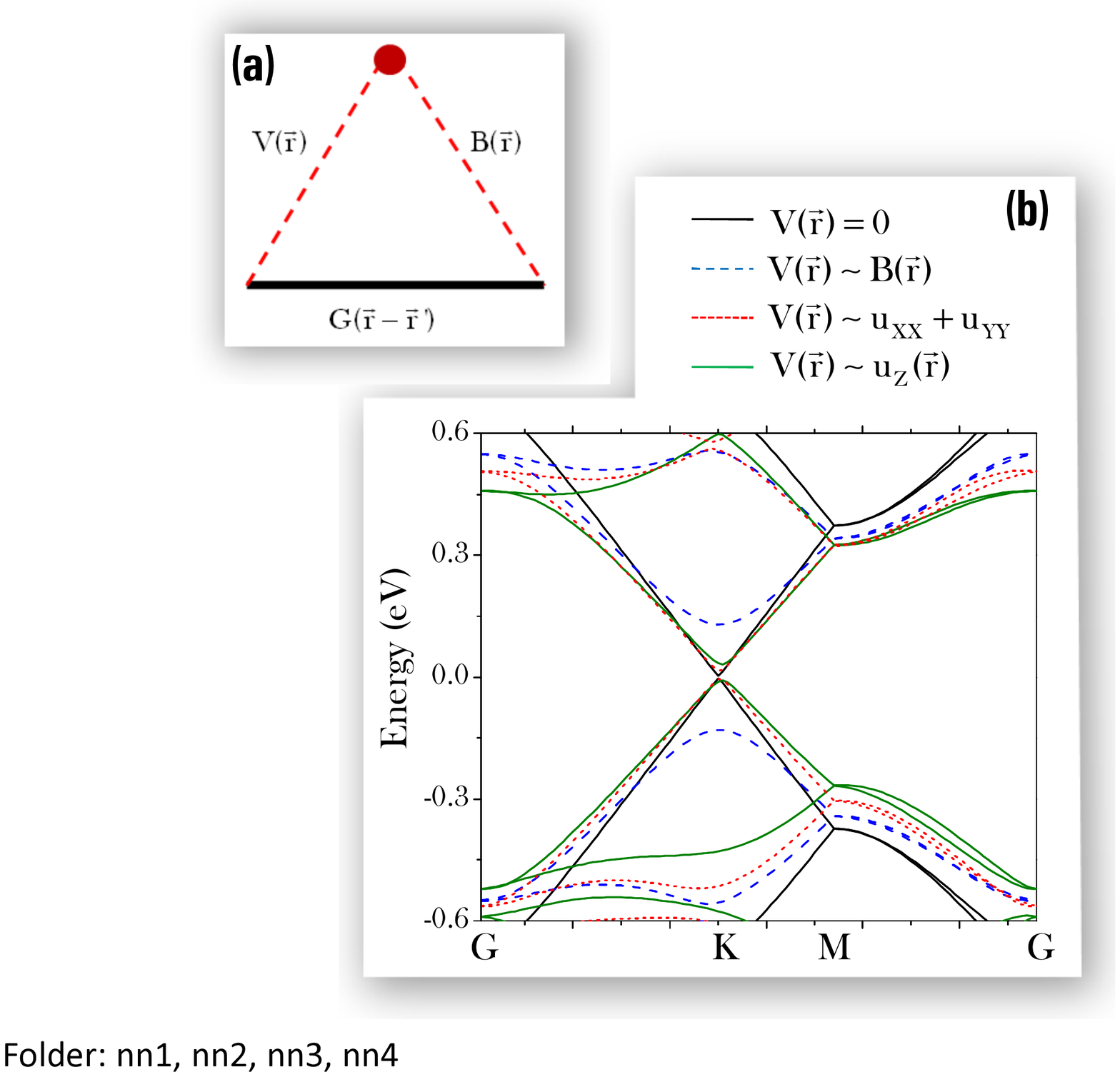}}
\caption[fig]{$\bold{(a)}$ Sketch of the diagram which describes correlations between the scalar potential and pseudomagnetic field, see Eq.~\ref{gap}. $\bold{(b)}$ Electronic bandstructure of graphene superlattice, where each supercell unit (as indicated in Fig. 1a) contains $40\times 40$ graphene lattice units. Out-of plane corrugations amplitude of $h_{0}=2\AA$ is used, leading to a non-homogenoeus pseudomagnetic field (see Fig. 1a) less than $100T$. Various scalar potentials as indicated are considered, where $V(\vec{r})=c_{0}B(\vec{r})$ and $V(\vec{r})=c_{0}u_{z}(\vec{r})$ leads to gap opening. $c_{0}$ is chosen such that max$[V(\vec{r})]=0.2V$.}
 \label{diagram}
\end{figure}

We calculate the second order diagram for the electron self energy
(effective potential) as shown in Fig.~\ref{diagram}a. At low
energies, $\omega=0$, and zero wave vector ${\bf k}=0$
(corresponding to the Dirac point), the form for the induced gap
reads
\begin{align}
\nonumber
\Delta &= - {\rm Tr} \left\{ \sigma_z \frac{2}{v_F} \int d^2 \vec{\bf k}
\frac{{\rm Im} \left( V_{-\vec{\bf k}} \right) \left[ ( \vec{\bf k}  \vec{\sigma} ) , (
\vec{\bf A}_{\vec{\bf k}} \vec{\sigma}   ) \right]}{ | \vec{\bf
k} |^2} \right\} \\
& \propto \int d^2 \vec{\bf k} \frac{{\rm Im} ( V_{-\vec{\bf k}} ) \left(
k_x A^y_{\vec{\bf k}} - k_y A^x_{\vec{\bf k}} \right)}{ | \vec{\bf
k} |^2} \label{gap}
\end{align}
where $\left[ \cdots \right]$ is the commutator, $V_{\vec{\bf k}}$
and ${\bf A}_{\vec{\bf k}}$ are Fourier components of scalar and
vector potentials, respectively. This equation shows that the gap
is induced through the correlations between the scalar potential
and the pseudomagnetic (synthetic magnetic) field, $B_{\vec{\bf
k}} = k_x A^y_{\vec{\bf k}} - k_y A^x_{\vec{\bf k}}$. We
characterize these correlations by the parameter $C$ such that
\begin{align}
\lim_{\vec{\bf k} \rightarrow 0}   ( B V )_{\vec{k}} &= C \label{cross}
 \end{align}
 or, alternatively,
 \begin{align}
 \lim_{| \vec{\bf r} - \vec{\bf r}' | \rightarrow \infty} \left\langle V ( \vec{\bf r} ) B ( \vec{\bf r}' ) \right\rangle &= C \delta^{(2)} ( \vec{\bf r} - \vec{\bf r}' ) \label{cross_1}
 \end{align}
The parameter $C$ has dimensions of energy. It is roughly given by
the value of the scalar potential times the number of flux quanta
due to the synthetic field over the region where the field and the
scalar potential are correlated. For $C \ne 0$, the integral in
Eq.~(\ref{gap}) diverges as $\vec{\bf k} \rightarrow 0$. Then, the
lower limit of the integral should be $k_{min} \approx \Delta /
v_F$, turning Eq.~(\ref{gap}) into a self consistent equation for
$\Delta$.

 \begin{figure*}
\scalebox{0.63}[0.63]{\includegraphics*[viewport=115 250 720 570]{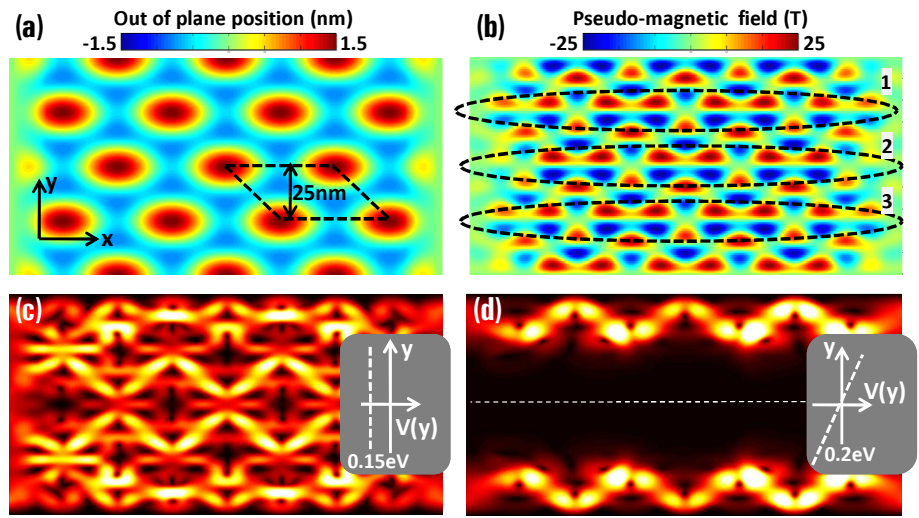}}
\caption[fig]{  $\bold{(a)}$ Finite size strain superlattice of dimensions $200nm\times 100nm$ (supercell as indicated) and its out-of-plane position. $\bold{(b)}$ The corresponding pseudomagnetic field generated. $\bold{(c)}$ Spatially resolved current density at Fermi energy $\epsilon_{f}=0$ and zero temperature, under infinitesimally small source-to-drain bias. Scalar potential is defined as $V(\vec{r})=\epsilon_{f}-\epsilon_{D}$, where $\epsilon_{D}$ is Dirac energy. Here, we assume $V(\vec{r})=0.15V$. $\bold{(d)}$ Same as previous, except now the scalar potential varies linearly along the width direction, i.e. $V(y)\propto y$, and $V=\pm0.2V$ along the two edges. Conducting channel $2$ is ``switched off'' with this type of scalar potential.}
 \label{superlattice}
\end{figure*}

In the diffusive regime, where electrons with momentum
$\vec{\bf k}$ have an elastic scattering time $\tau_{\vec{k}}$,
the divergence in the integral in Eq.~(\ref{gap}) has to be cutoff
at a momentum $k_{min}$ such that $v_F k_{min} \approx
\tau_{k_{min}}^{-1}$. For a periodic superlattice, the integral in
Eq.~(\ref{gap}) has to be replaced by a sum over reciprocal
lattice vectors, $\vec{\bf G}$. For graphene in the diffusive
regime, resonant scatterers\cite{SPG07,WYLGK10} or substrate
charges\cite{NM07,AHGS07} give rise to a dependence
$\tau^{-1}_{\vec{\bf k}} \propto n_i | \vec{\bf k} |^{-1}$, where
$n_i$ is the concentration of scatterers, so that the lower cutoff
in Eq.~(\ref{gap}) is $k_{min} \propto \sqrt{n_i}$. Using this
cutoff, we can write
 \begin{align}
 \Delta &\approx C \, \, \, \log \left[ \frac{1}{{\rm Max} ( k_{min} , \Delta / v_F ) \, \, a} \right]
 \end{align}
where $a \approx 1.4\,\AA$ is the distance between nearest carbon
atoms, which sets the scale of the high momentum limit in
Eq.~(\ref{gap}). Next, we extend
this general argument to specific examples.

\section{Possible physical realizations}

\subsection{Global gap in strain superlattices}

We show results for the gap opening
due to correlation between the synthetic magnetic field and a
scalar potential in strain graphene superlattice.
We assume that the graphene layer is corrugated, as depicted in
Fig.~\ref{sketch}a. Depending on the lattice mismatch with the
underlying substrate, the supercell size could vary from
$\approx 10\times10$ (e.g. iridium) to $\approx 50\times50$ (e.g. boron nitride)
times that of graphene unit cell. Here, we assumed supercell to be $40\times40$.
The in-plane displacements were relaxed in order to minimize the
elastic energy. Details of the implementation
based on continuum elasticity model are given in Appendix A.

\begin{figure*}
\scalebox{0.5}[0.5]{\includegraphics*[viewport=40 180 720 570]{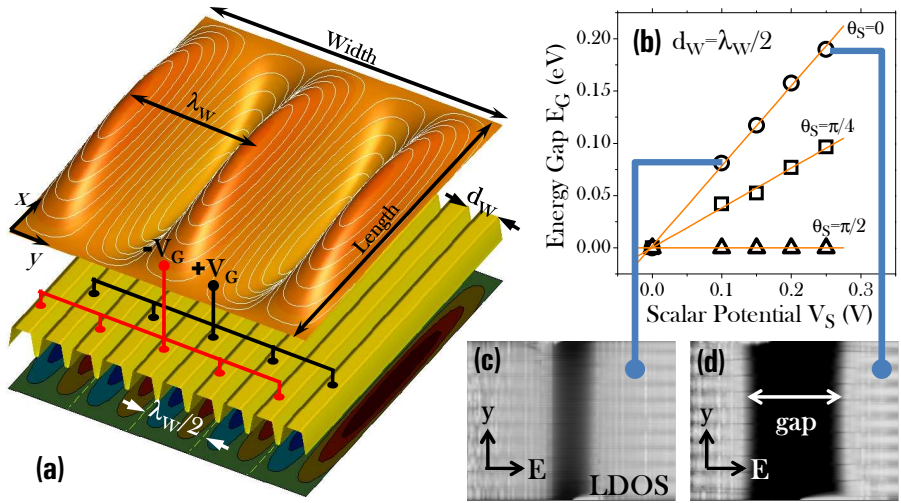}}
\caption[fig]{ $\bold{(a)}$ Schematic of the elongated and wrinkled graphene device, of dimensions $100nm\times 100nm$. The transport is along $x$ direction. In this calculation, the quasi-$1D$ ripple along $y$ is assumed to have a wavelength of $\lambda_{w}=20nm$. The gates are aligned underneath the graphene, with alternating potentials of $\pm V_{G}$ and spaced at distance $d_{w}=10nm$ (same period as the underlying pseudomagnetic field). $\bold{(b)}$ The gates results in an approximately sinusoidally varying scalar potential in the graphene, with amplitude $V_{s}$. Energy gaps as a function of $V_{s}$ is shown, for different amount of phase shift between the gates and underlying pseudomagnetic field, $\theta_{s}=0,\tfrac{\pi}{4},\tfrac{\pi}{2}$. $\bold{(c)}$ and $\bold{(d)}$ shows the calculated local-density-of-states as function of $y$ and energy $E$, along an $x$-cut in the middle of the device.}
 \label{wrinkle}
\end{figure*}

After minimizing the elastic energy, the resultant strains lead to an
 underlying non-homogeneous pseudomagnetic field as depicted in  Fig.~\ref{sketch}a.
Fig.~\ref{diagram}b shows the calculated electronic bandstructure
of the superlattice, after considering various types of scalar
potentials. Strain alone, $V(\vec{\bf r})=0$, does not produce a
gap. Although strains can induced a scalar potential of type
$V(\vec{\bf r})\propto u_{xx}+u_{yy}$, the correlation between
this potential and the synthetic magnetic field due to the same
strains is also zero. On the other hand, a scalar potential
proportional to the height corrugation, i.e. $V(\vec{\bf
r})\propto u_{z}$, leads to the appearance of a gap, albeit a
small one. Such a scalar potential could be induced by the
substrate through an existence of electric field perpendicular to
the graphene layer~\cite{GL10}.  Graphene superlattices induced by
commensuration effects between the mismatch in the lattice
constants of graphene and the
substrate\cite{Zetal07,Vetal08,Metal08,Petal08} will, in general
lead to the effect considered here. The existence of a gap at the
Dirac energy in strained graphene superlattice is also consistent
with observations which show gaps in very clean samples which are
commensurate with the substrate\cite{Zetal07,LLA09,HK10}.

In essence, strains generally induce both scalar and vector
potentials. However, cross correlations between the scalar
potential and the synthetic magnetic field, as in
Eq.~(\ref{cross_1}), vanish in many cases. In particular, the gap
should be zero if the system remains symmetric with respect to
inversion\cite{noteMGV07}. A significantly large gap is obtained
when the scalar potential and pseudomagnetic field are perfectly
correlated, as shown in Fig.~\ref{diagram}b for the case when
$V(\vec{\bf r})\propto B(\vec{\bf r})$. This observation is
consistent with the general arguments that we presented in
Sec.~\ref{gen}. See also Appendix A, where a general expression
for $\Delta$ in strain superlattice is derived.

Next, we examine a slightly different scenario, where electrostatic gates
are used to engineer the scalar potential so as to realise the correlation with
the underlying strain-induced pseudo magnetic field.

\subsection{Local gap in strain superlattices through quantum transport calculations}

Here, we examine numerically the effect a local electric scalar potential on the electronic transport properties of strained graphene superlattices.
The Hamiltonian accounting for nearest neighbor interactions between $p_{z}$ orbitals is given by\cite{W47},
\begin{align}
{\cal H} = \sum_{i}V_{i}a_{i}^{\dagger}a_{i} + \sum_{ij}t_{ij}a_{i}^{\dagger}a_{j}
\label{hamilTB}
\end{align}
where $V_{i}$ is the on-site energy due to the scalar potential
$V(\vec{\bf r})$ and $t_{ij}=t(1-\tfrac{\beta}{a}(a_{ij}-a))$ is
the hopping energy. $a_{ij}$ is the new bond length after strain.
To facilitate the application of various numerical techniques, the
problem is partitioned into block slices as shown in
Fig.~\ref{qt}. The retarded Green's function in $\Omega_{0}$, the
device region of interest, can then be written as (see
\cite{DV08,D97,HJ10} for general theory),
\begin{align}
{\cal G} = \left(\epsilon_{f}{\cal I}-{\cal H}_{0}-\Sigma_{L}-\Sigma_{R}\right)^{-1}\equiv A^{-1}
\label{greenrt}
\end{align}
where $\epsilon_{f}$ is the Fermi energy, and $\Sigma_{L/R}$ are defined as $\Sigma_{L}=\tau^{\dagger}g_{L}\tau$ and $\Sigma_{R}=\tau g_{R}\tau^{\dagger}$ respectively. $g_{L/R}$ are the surface Green's function, which can be obtained numerically through an iterative scheme \cite{SSSR85} based on the decimation technique (see e.g. \cite{GTFL83}).  Various physical quantities of interest such as the transmission, current/charge density, local density-of-states can be obtained once ${\cal G}$ is determined. See Appendix B for a more detailed description of the numerics.

We consider a finite size strain superlattice of dimension $200nm\times100nm$, as depicted in  Fig.~\ref{superlattice}a. The corresponding pseudomagnetic field is shown in Fig.~\ref{superlattice}b. Transport in non-homogeneous magnetic field
is dominated by bulk ``magnetic'' states known as snake states\cite{MDE08}.
Snake states has been observed in high mobility 2D electron gas system, through controlled engineering of magnetic field via lithographic patterning of ferromagnetic or superconducting thin films\cite{ye95,nogaret00,carmona95}.

Fig.~\ref{superlattice}c plots the current density due to current injection from the left contact, biased at Fermi energy of $150meV$.
As depicted, current flows in regions where $B(\vec{\bf r})\approx 0$, along the direction $\pm\nabla B(\vec{\bf r})\times \hat{z}$. Unlike the case of a real magnetic field, these snake states are non-chiral, with forward and backwards going states residing in opposite valleys. In the absence of short range scatterers, these states are relatively protected. Fig.~\ref{superlattice}b shows three conducting snake channels which forms the backbone for the conduction. Applying the general principle described in Sec.~\ref{gen}, we apply a scalar potential that approximately correlates with the pseudomagnetic field of the middle channel to open up a local gap. Indeed, a local gap is opened, impeding current flow along this channel, as shown in Fig.~\ref{superlattice}d. Such a scalar potential can be realised experimentally with electrostatic side gates.
This effect could be exploited in current guiding devices \cite{williams11}.

\subsection{Opening gaps in suspended graphene via wrinkles}

Wrinkles are common feature in very clean suspended graphene
samples, leading to finite strains. Partial control of these
wrinkles can be achieved by adjusting the temperature, as in some
cases, they are induced by the mismatch in thermal expansion
coefficients between graphene and the
substrate\cite{Betal09,wang11}. We assume that the deformation is
described by the profile proposed in~\cite{CM03}, as illustrated
in Fig.~\ref{sketch}a. The resulting synthetic field is discussed
elsewhere~\cite{GHL09}. Fig.~\ref{sketch}b illustrates the
accompanied pseudo-magnetic field.

A sinusoidal-like scalar potential within graphene is induced by
gates shown in Fig.~\ref{wrinkle}a, which are tailored to
correlate with the synthetic magnetic field induced by the
strains. The results in Fig.~\ref{wrinkle}b show that a gap is
generated, whose magnitude is proportional to the scalar
potential. We can estimate the gap induced by a combination of
strains which are changed by $\delta u$ over an area of spatial
scale $\ell$ and a scalar potential of value $\delta V$ on a
region of the same size. The synthetic magnetic field is of order
$B \sim ( \beta \delta u ) / ( a \ell )$. Then
 \begin{align}
 \Delta \sim C \approx \beta \delta u \delta V \frac{\ell}{a}
 \label{gap12}
 \end{align}
Eq.~(\ref{gap12}) is consistent with numerical results obtained in
Fig.~\ref{wrinkle}b. Fig.~\ref{wrinkle}c and d clearly show the
global nature of the gap generated. Effectively, the gate
controlled gap allows the device in Fig.~\ref{wrinkle} to be
operated as a graphene transistor.

 Next, we consider situation where correlation is less than perfect.
 From Eq.~(\ref{gap12}), we note that even for small variations in the strain, $\delta u \ll 1$, the gap
can be of order of the potential fluctuations, $\delta V$, if the
correlations between the scalar potential and the synthetic field
are maintained over long distances, $\ell \gg a$.
Fig.~\ref{wrinkle}b considers the case where there is a phase
shift between the pseudomagnetic field and scalar potential. This
corresponds to a decreasing $\ell$ in Eq.~(\ref{gap12}). Indeed
the gap reduces as expected.  In general, the presence of a gap is
robust against reasonable degree of local disorder, since
inversion symmetry is still absent in most part, see also Eq.\,(4)
and related discussions.

\section{Discussions}

In this section we discuss several issues related with the
previous consideration and the ways of further development.

\subsection{Renormalization of the Fermi velocity}
As evident from Eq.~(\ref{sigma}), there are self energy
corrections due to quadratic terms in the scalar and vector
potentials. These terms lead to logarithmic corrections in the
Fermi velocity via the renormalization of the residue of the Green
function:
\begin{equation}
\left. \frac{\partial \Sigma \left( \vec{\bf k},E\right) }{\partial E} \right|_{E=0} =-%
 \frac{\ln \Lambda }{2\pi v_F^2}\left( V_{\vec{\bf k}}V_{-{\vec\bf
k}}+\vec{\bf A}_{\vec{\bf k}}\vec{\bf A}_{-\vec{\bf k}}\right)
\end{equation}
where $\Lambda \gg \left| \vec{\bf k} \right|$ is a momentum cutoff.

This renormalization has been previously found in non-linear sigma
models\cite{F86,L93,NTW94,Z98}. These logarithmic corrections also
influence the kinetic equation which describes transport
processes\cite{AK07}, in these terms they describe a pseudo-Kondo
effect due to interband scattering (in the Dirac point, the
energies of electron and hole states coincide which provides a
necessary degeneracy). Due to these corrections, the Fermi
velocity decreases in the presence of scalar and gauge disorder.
When this effect is studied simultaneously with the increase
induced by the Coulomb interaction non trivial new phases can
arise\cite{SGV05,HJV08,FA08}.

\subsection{Gauge field due to topological defects}
The sublattice and valley symmetries of graphene allow for the
definition of a second gauge field, which hybridizes states from
different valleys, and does not commute with the intravalley gauge
field due to long wavelength strains\cite{NGPNG09,VKG10}. This
field can be induced by topological defects, such as heptagons and
pentagons. These defects are present at dislocations and grain
boundaries, and they can be ordered periodically forming
superlattices. If the synthetic magnetic field associated with
this field is correlated with a scalar potential, a gap inducing
term is generated.

The gauge field due to topological defects has the form
\begin{align}
{\cal H}_{\tilde{A}} &= -  v_F \left[ \tau_x \tilde{A}_y ( \vec{\bf r} ) + \tau_y \sigma_z \tilde{A}_y ( \vec{\bf r} ) \right]
\end{align}
By using perturbation theory, as in Eq.~(\ref{sigma}), we obtain a
self energy which is proportional to the cross correlations
between the scalar potential and the gauge field, multiplied by
the operator $\tau_y \sigma_x$. Modulated Zeeman couplings can also lead
to synthetic fields which act on the spin, allowing for the
possibility of spin gaps as well.

\subsection{Interplay with magnetic field}

 The gap studied here is defined
in the whole sample, although its value should be roughly
inversely proportional to the ratio between the total area and the
area where the synthetic magnetic field and scalar potential are
correlated. The sign of the gap is determined by the scalar
potential. Localized states will be formed at boundaries between
regions where the gaps have different signs, similar to the edge
states in topological insulators \cite{QZ10,M10}.

A periodic magnetic field, when correlated with a scalar potential
leads to a gap whose signs are opposite in the two
valleys\cite{S09}. A combination of this gap and the gap due to
strains leads to gaps of different values in the two valleys,
allowing for the control of the valley and sublattice degrees of
freedom. For example, combined strain and synthetic magnetic field
could be useful for valleytronics\cite{LG10}. The realization of
other synthetic fields might open new functionalities for graphene
that cannot be achieved with other materials.

\section{Conclusion}
We have discussed a novel way in which a combination of long
wavelength strains and a long wavelength correlated potential can
lead to a gap in the electronic spectrum of graphene. Such
situation can occur naturally, because of correlations between a
periodic substrate and graphene, or it can be engineered in a
controlled way using electrostatic gates. Since the effect is induced by long wavelength,
smooth perturbations, a gap can be induced without increasing the
amount of scattering in the system. Finally, valley polarized edge
states will be generated, as the band structure of the modified
system resembles the spectrum of a quantum Hall insulator.

\section{Acknowledgements}
TL acknowledges funding from INDEX/NSF
(US). FG acknowledges financial support from MICINN (Spain) through grants
FIS2008-00124 and CONSOLIDER CSD2007-00010, and
from the Comunidad de Madrid, through NANOBIOMAG. The work of MIK is part of
the research program of the Stichting voor Fundamenteel
Onderzoek der Materie (FOM), which is financially supported by the Nederlandse Organisatie voor Wetenschappelijk Onderzoek (NWO).
Computational resources is provided by Network for Computational Nanotechnology (NCN) at Purdue University.

\appendix

\section{\label{gapexpress} Expression for energy gap in superlattice}
Strains induce scalar and gauge potentials\cite{VKG10}. We study the correlations between these potentials when the strains are induced by modulations in the vertical displacement of the layer, $h ( \vec{\bf r} )$. We assume that the in plane displacements relax in order to minimize the elastic energy. The strains are\cite{GHL08}
\begin{align}
u_{ij} ( \vec{\bf k} ) &= \frac{\lambda + \mu}{\lambda + 2 \mu} \frac{k_i k_j \left[ k_x^2 h_{yy} ( \vec{\bf k} )
+ k_y^2 h_{xx} ( \vec{\bf k} ) - 2 k_x k_y h_{xy} ( \vec{\bf k} ) \right]}{| \vec{\bf k} |^4}
\label{strains}
\end{align}
where $h_{ij} ( \vec{\bf k} )$ are the Fourier transforms of the tensor
\begin{align}
h_{ij} ( \vec{\bf x} ) &= \partial_i h \partial_j h
\end{align}
In terms of the strain tensor, the scalar and vector potentials are\cite{SA02b,M07,VKG10}:
\begin{align}
V ( \vec{\bf r} ) &= g \left[ u_{xx} ( \vec{\bf r} ) + u_{yy} ( \vec{\bf r} ) \right] \nonumber \\
A_x ( \vec{\bf r} ) &= \frac{\beta}{a} \left[ u_{xx} ( \vec{\bf r} ) - u_{yy} ( \vec{\bf r} ) \right] \nonumber \\
A_y ( \vec{\bf r} ) &= 2 \frac{\beta}{a} u_{xy} ( \vec{\bf r} )
\label{fields}
\end{align}

Using Eq.~(\ref{gap}) of the main text, we obtain
\begin{align}
\Delta &= \frac{g \beta}{a} \frac{(\lambda + \mu )^2}{(\lambda + 2 \mu )^2} \times \nonumber \\
&\times \int d^2 \vec{\bf k} \frac{\left| k_x^2 h_{yy} ( \vec{\bf k} ) + k_y^2 h_{xx} ( \vec{\bf k} )
- 2 k_x k_y h_y ( \vec{\bf k} ) \right|^2 \cos ( 3 \theta_{\vec{\bf k}} )}{\left| \vec{\bf k} \right|^4}
\end{align}
This expression is zero, as $\theta_{- \vec{\bf k}} = \theta_{\vec{\bf k}} + \pi$. While the scalar and
gauge potentials are correlated, their correlation does not contribute to the formation of a
global gap.

\section{Quantum transport methods.}
\begin{figure}
\scalebox{0.45}[0.45]{\includegraphics*[viewport=110 170 720 490]{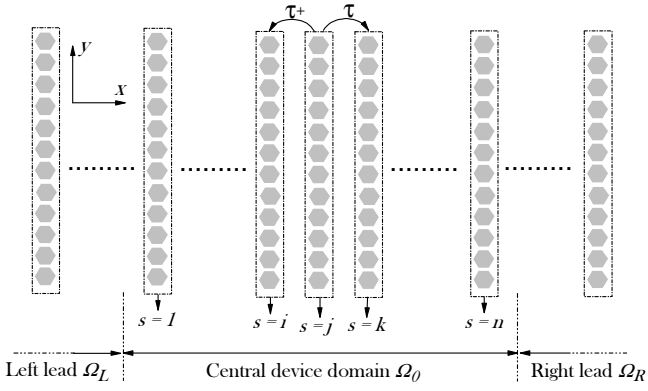}}
\caption[fig]{\bfseries Numerical approach: \normalfont The graphene ribbon is partitioned into block slices along the $x$-direction (transport) as indicated. Lattice interactions within each block is described by $\alpha$. Nearest neighbor blocks interactions are represented by $\tau$. Device domain $\Omega_{0}$ will include the strains and scalar potential $V(\vec{r})$. Left/right leads regions ($\Omega_{L/R}$) are assumed unstrained and electrically doped, due to charge transfer from contacts.   }
 \label{qt}
\end{figure}

The Hamiltonian accounting for nearest neighbor interactions between $p_{z}$ orbitals is given by\cite{W47},
\begin{align}
{\cal H} = \sum_{i}V_{i}a_{i}^{\dagger}a_{i} + \sum_{ij}t_{ij}a_{i}^{\dagger}a_{j}
\label{hamilTB}
\end{align}
To facilitate the application of various numerical techniques, the problem is partitioned into block slices as shown in Fig. \ref{qt}. The retarded Green's function in $\Omega_{0}$, the device region of interest, can then be written as (see \cite{DV08,D97,HJ10} for general theory),
\begin{align}
{\cal G} = \left(\epsilon_{f}{\cal I}-{\cal H}_{0}-\Sigma_{L}-\Sigma_{R}\right)^{-1}\equiv A^{-1}
\label{greenrt}
\end{align}
where $\epsilon_{f}$ is the Fermi energy, and $\Sigma_{L/R}$ are defined as $\Sigma_{L}=\tau^{\dagger}g_{L}\tau$ and $\Sigma_{R}=\tau g_{R}\tau^{\dagger}$ respectively. $g_{L/R}$ are the surface Green's function, which can be obtained numerically through an iterative scheme \cite{SSSR85} based on the decimation technique (see e.g. \cite{GTFL83}). It is also useful to define the quantity, broadening function, $\Gamma_{L/R}\equiv i(\Sigma_{L/R}-\Sigma_{L/R}^{\dagger})$. Physical quantities of interest such as the transmission ${\cal T}$ is given by,
\begin{align}
{\cal T}=\mbox{Tr}\left([\Gamma_{L}]^{1}_{1}[{\cal G}]^{1}_{n}[\Gamma_{R}]^{n}_{n}[{\cal G}^{\dagger}]^{n}_{1}\right)
\label{trandef}
\end{align}
Energy gaps, as seen in Fig. 4b of main manuscript, is estimated by the onset of increase in ${\cal T}$.
The electron density $n(\vec{r})$ at slice $j$ is obtained from the diagonals elements of ${\cal G}^{n}$, given by,
\begin{align}
[{\cal G}^{n}]^{j}_{j}=f_{L}[{\cal G}]^{j}_{1}[\Gamma_{L}]^{1}_{1}[{\cal G}^{\dagger}]^{1}_{j}+f_{R}[{\cal G}]^{j}_{n}[\Gamma_{R}]^{n}_{n}[{\cal G}^{\dagger}]^{n}_{j}
\label{gndef}
\end{align}
Local density-of-states (as seen in Fig. 4c-d of main manuscript) is obtained from Eq.~\ref{gndef} by simply setting $f_{L}=f_{R}=1$.
Current density $j(\vec{r})$ (as seen in Fig. 3c-d of main manuscript), flowing from slice $j$ to $j+1$ is given by the diagonal of $J$, given by,
\begin{align}
[J]^{j}_{j+1}=\tfrac{2q}{h}\left([A]^{j}_{j+1}[{\cal G}^{n}]^{j+1}_{j}-[A]^{j+1}_{j}[{\cal G}^{n}]^{j}_{j+1}\right)
\label{jcdef}
\end{align}
where,
\begin{align}
\nonumber
[{\cal G}^{n}]^{j+1}_{j}=f_{L}[{\cal G}]^{j+1}_{1}[\Gamma_{L}]^{1}_{1}[{\cal G}^{\dagger}]^{1}_{j}+f_{R}[{\cal G}]^{j+1}_{n}[\Gamma_{R}]^{n}_{n}[{\cal G}^{\dagger}]^{n}_{j}\\
[{\cal G}^{n}]^{j}_{j+1}=f_{L}[{\cal G}]^{j}_{1}[\Gamma_{L}]^{1}_{1}[{\cal G}^{\dagger}]^{1}_{j+1}+f_{R}[{\cal G}]^{j}_{n}[\Gamma_{R}]^{n}_{n}[{\cal G}^{\dagger}]^{n}_{j+1}
\label{gndef2}
\end{align}
As apparent from Eq.~(\ref{trandef})-(\ref{gndef2}), it is not neccessary to obtain the full matrix ${\cal G}$. Through commonly used recursive formula of the Green's function derived from the Dyson equation and the decimation technique, one could obtain these block elements of the Green's function, $[{\cal G}]^{i}_{j}$, in a computationally/memory efficient manner. Details of this numerical recipe are described elsewhere\cite{LA09}.


\end{document}